\documentclass{scrartcl}
\usepackage{amsmath}
\usepackage{graphicx}
\def\lsim{\mbox{\raisebox{-.6ex}{~$\stackrel{<}{\sim}$~}}}

\title{Chain Inflation Reconsidered}
\author{    James M. Cline \thanks{jcline@hep.physics.mcgill.ca},
Guy D. Moore \thanks{guymoore@hep.physics.mcgill.ca},
  Yi Wang \thanks{wangyi@hep.physics.mcgill.ca}
  \\
  \textit{Physics Department, McGill University, Montreal, H3A2T8,
    Canada}
}
\date{}

\begin{document}
\maketitle

\begin{abstract}
  We investigate density perturbations in the chain inflation scenario,
  where the inflaton undergoes successive tunneling transitions along
  one field direction.  First we show that when the bubble walls
  associated with such a phase transition meet, they induce the next
  phase transition and continue to propagate as phase interfaces.  Then
  we present an analytical calculation of the density fluctuations and
  an estimate of non-Gaussianities such as $f_{NL}$, which we find to be
  small (of order 1).  To get the right amplitude for the power
  spectrum, there have to be $10^8$ phase transitions per Hubble time, a
  significant model building challenge.  We find that working models
  of chain inflation must be rather strongly coupled, and thus have a
  very limited range of validity as effective field theories. 
  We discuss generalizations to the multiple field case, the curvaton
  scenario, and noncanonical kinetic terms. 
\end{abstract}

\section{Introduction}

In a scenario dubbed ``chain inflation'' \cite{Freese:2004vs,
  Freese:2005kt} it was suggested that the dynamics of inflation may be dominated by a
series (chain) of phase transitions, each induced by the quantum
nucleation of bubbles (tunneling events).  Provided that each phase
transition occurs much less than one Hubble time after the previous one,
this scenario is free from the problem of ``old inflation''
\cite{old_inflate} that the space still in the previous phase continues
to inflate and prevents the completion of a phase transition.  The need
for several phase transitions per Hubble time, and the requirement that
inflation persist for at least $\sim 60$ Hubble times to solve the
flatness and horizon problems, means that the scenario
requires a potential with a large number of minima, tunneled through in
sequence.  This is a challenge for model building but not an
insurmountable one.

The density perturbations arising from chain inflation have been
investigated in \cite{Feldstein:2006hm, Chialva:2008zw,
Chialva:2008xh}. A simplified model of chain inflation was proposed
in \cite{Huang:2007ek}. 
However, in these works either the density fluctuation has to be
determined numerically, or features of standard slow roll inflation
(with vacuum fluctuations instead of fluctuations from tunneling) are
assumed in the calculation. Thus an analytical calculation for the
perturbation theory is still absent for chain inflation.

In this paper we demonstrate three properties of chain inflation which seem
not to have been addressed, and which are each troubling from the point of
view of building realistic models.  First, when the chain of tunneling
events takes place in a single field direction, the collision of two bubble
walls will trigger a nucleation of the next phase.  Second, the rate at
which the field, responsible for the chain of phase transitions, moves along
its potential is controlled by the rate of bubble nucleations.  But since
bubble nucleations are stochastic in nature, the fluctuations in the field
value, which become curvature perturbations, are controlled by the Poisson
statistics of nucleation events.  To meet current observational limits on
curvature perturbations requires of order $10^8$ phase transitions per
Hubble time during the observable part of inflation, requiring a potential
with of order $10^{10}$ or more local minima to tunnel through
successively.  The third problem is that it is challenging to provide a
potential where the nucleation rate between minima is large enough to
support chain inflation, in a field theory that remains valid to scales 
that are much higher than the mass of the field.  

Our paper is organized as follows: in Section
\ref{sec:setup-backgr-dynam}, we set up our model and discuss the
dynamics of bubble collisions which are essential to understanding
the overall evolution of the inflaton.
In section \ref{sec:evolution} we compute the time dependence of the mean field
which constitutes the homogeneous and isotropic background. In Section
\ref{sec:power-spectrum}, we consider the density perturbations and
we calculate the power spectrum
analytically, showing that it takes a simple form. 
Section \ref{sec:models} investigates the simplest explicit realization of a chain
inflation model and identifies a narrow region of parameter space in which
it satisfies observational and internal consistency requirements.
In Section
\ref{sec:non-gaussianities}, we consider three- and higher-point
correlations. We consider generalizations in Section
\ref{sec:generalizations} and conclude in Section
\ref{sec:conclusion}.

\section{Setup and bubble collision dynamics}
\label{sec:setup-backgr-dynam}

\begin{figure}[hptb]
\centering
\includegraphics[width=0.5 \textwidth]{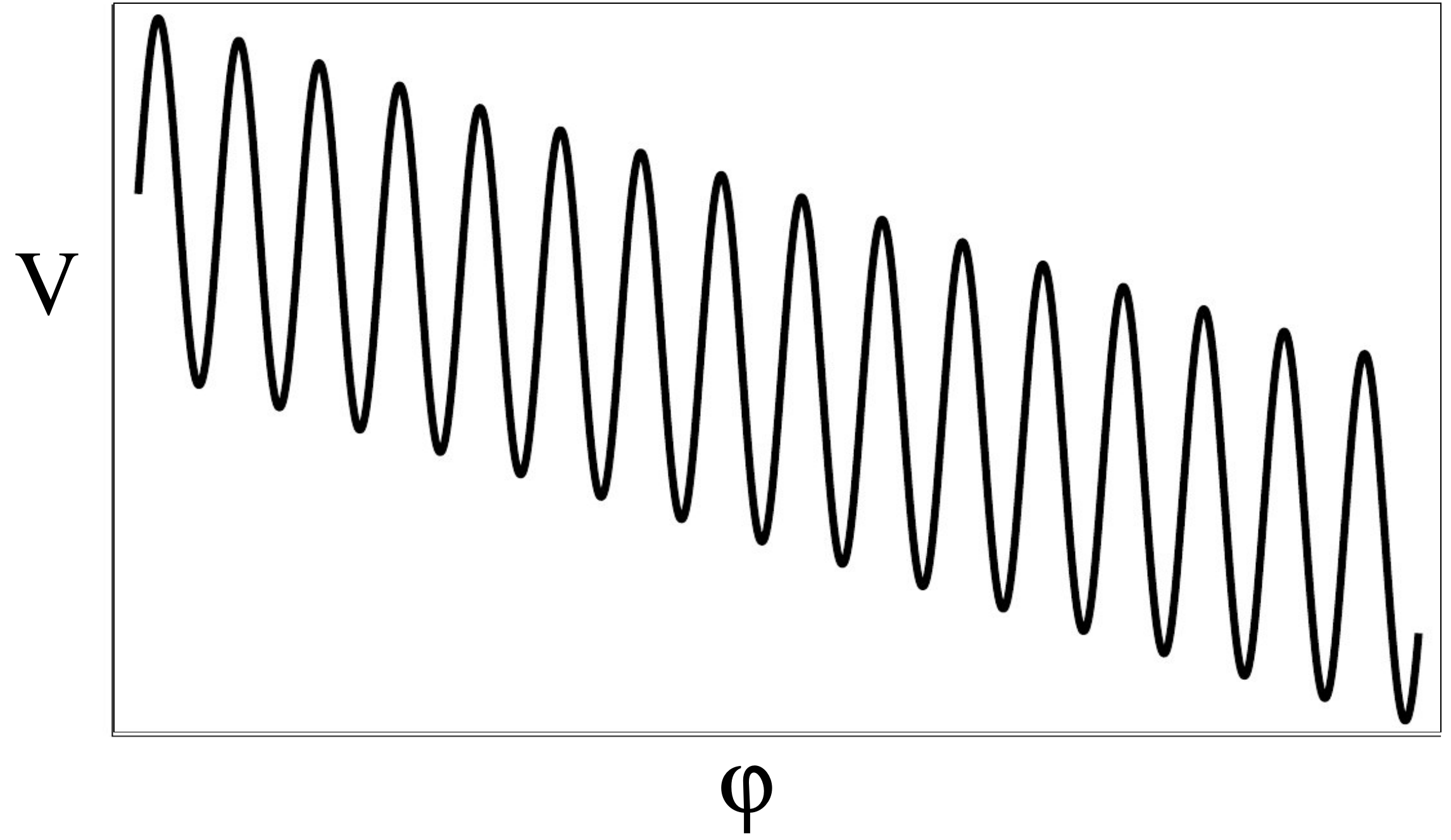}
\caption{\label{fig:potential} A sample potential for single-field
  chain inflation, as in equation (\ref{eq:sample-potential}). }
\end{figure}

We consider a simple model of chain inflation, with a single scalar
field whose potential
is a combination of an oscillation and a slope.  Specifically, we
consider
\begin{align}\label{eq:sample-potential}
  V(\varphi) = V_0(\varphi)+V_1 \sin(\omega\varphi)~,
\end{align}
where $V_0$ is a slowly varying potential and $V_1$, $\omega$ are
constants. A potential of this kind is illustrated in figure
\ref{fig:potential}.  In practice the pertinent features of the
potential are that each minimum is lower than the last, and that the
nucleation rate between each pair of minima is approximately the same.

We will also assume that the mean spacetime separation between
nucleation events is large compared to the size of a bubble at
nucleation.  Otherwise, a description in terms of bubble nucleations
does not make sense.  Since nucleation events must in turn be common on
the Hubble scale, this ensures that the instanton method
\cite{Coleman:1977py, Coleman:1980aw}  applies.  
We will also assume
that most of the energy from the potential drop between minima goes into
accelerating the bubble wall.

Unlike the case in \cite{Coleman:1977py, Coleman:1980aw},
there is a chain of metastable vacua. As we now argue, in this case
collisions of bubble walls  trigger the subsequent phase nucleation.
(This phenomenon has been previously discussed in ref.\ \cite{Easther:2009ft}.)
Specifically, numbering the potential minima from high to low as 
$\ldots,1,2,3,\ldots$, when an interface between minima 1 and 2 meets
another interface between 1 and 2, they pass through each other and
become interfaces between minima 2 and 3.
To see why, note first that the walls become very relativistic and
Lorentz contract to much narrower than the ``natural'' width set by the
mass (second derivative of the potential).  The process of walls
crossing therefore occurs on length and time scales shorter than the
natural scale in the potential.  Therefore it is controlled by the
kinetic term.  Scalar field dynamics governed by only the kinetic term is
free field theory, which is solved by the linear wave equation.
According to the linear wave equation, when two
interfaces meet, in which the field changes from value $\varphi_0$ to value
$\varphi_0 + 2\pi/\omega$,  they pass through each other, turning into
interfaces where $\varphi$ changes from $\varphi_0 + 2\pi/\omega$ to
$\varphi_0 + 4\pi/\omega$.  This is precisely the same behavior as colliding
kinks in the sine-Gordon model.

\begin{figure}[t]
\centering
\includegraphics[width=0.6\textwidth]{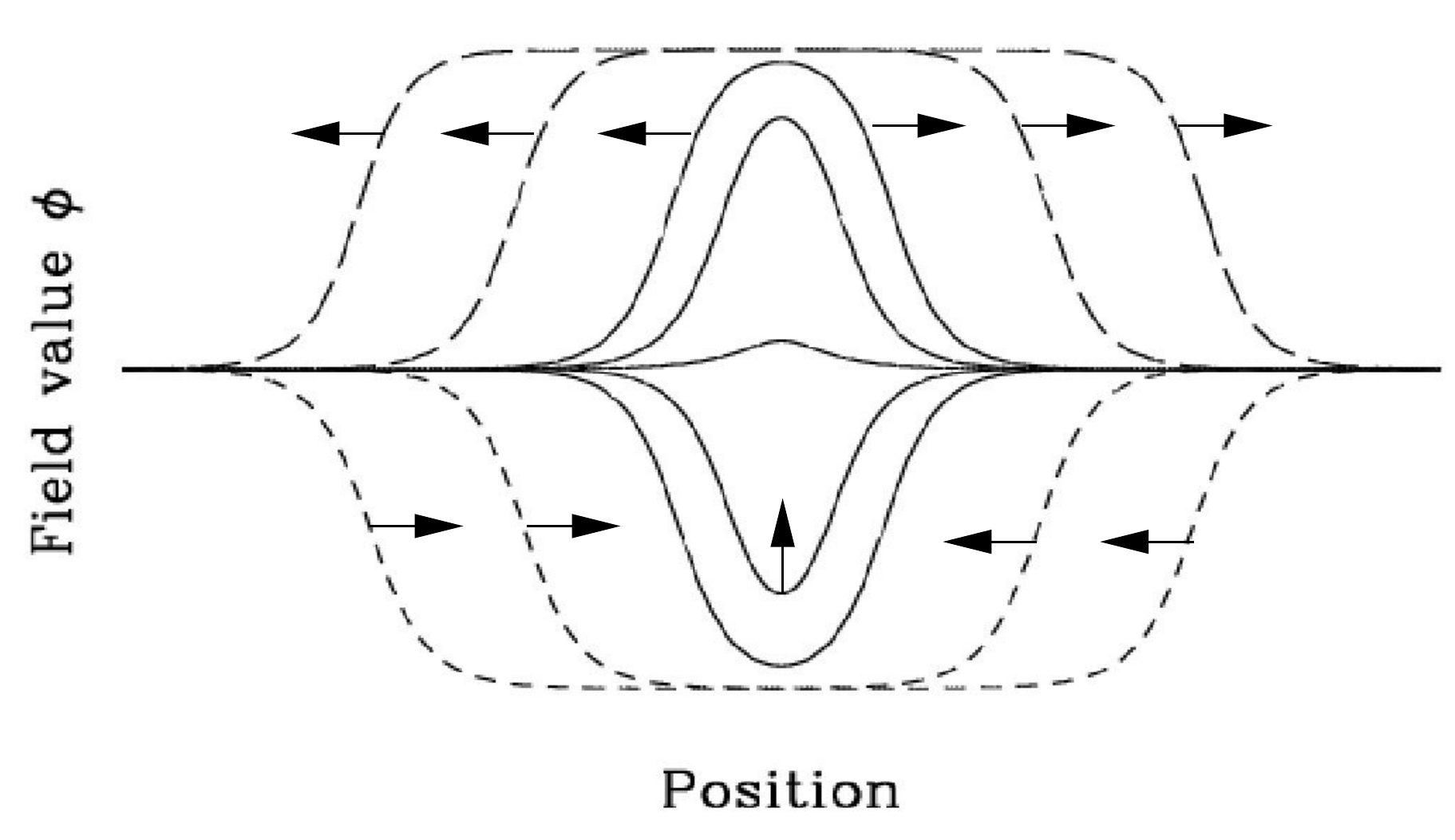}
\caption{\label{fig:collide}Collision of bubble walls.  Short-dashed
  lines show $\varphi$ at two moments before the walls meet; solid lines
  show $\varphi$ at a series of times as the walls meet; long dashes show
  two times after the walls have crossed.  When the walls cross, they
  switch from connecting minima 1 and 2, to connecting minima 2 and 3.
  }
\end{figure}

These remarks apply both in our case, and to
the case of a potential with only two minima.  When there are only two
minima, the region between the bubble walls is displaced from the
potential minimum and the field there oscillates, with a time scale
set by the parameters of the potential.  These oscillations exhaust the
energy of the bubble walls, which slow and stop, completing the
transition.  But in our case, the new value
of the field after the walls have crossed is again a minimum of the
potential, and the field stays there.  After passing each other, the
walls are again interfaces connecting a higher to a lower minimum, and
they continue to accelerate.  We have verified these
statements by simulating wave interactions in 1+1 dimensional classical
field theory using the potential from Eq.~(\ref{eq:sample-potential}).
Some snapshots of the field configuration as two walls meet are provided
in Figure \ref{fig:collide} as an illustration.

Since the bubble walls do not stop or dissipate their energy when they
collide, no thermal radiation is created by bubble
collisions. Thus we do not need to worry about the perturbations and
back-reaction from the radiation components created by the wall
collisions.  Since energy in bubble walls redshifts like
radiation,\footnote{this can be most easily established by
writing the stress-energy tensor for a bubble wall with tension
$\epsilon$ at rest, $T_{\mu\nu} = {\rm
diag}[\epsilon,0,-\epsilon,-\epsilon]$  and then
boosting by a large factor for a bubble moving near the speed of
light.  The diagonal part after boosting is $T_{\mu\nu} = 
\epsilon\gamma^2{\rm diag}[1,v^2,-\gamma^{-2},-\gamma^{-2}]$
from which it follows that the pressure is $p = 
\frac13(\gamma^2 v^2-2)\epsilon \cong \gamma^2\epsilon/3 = \rho/3$.}
they also do not come to dominate the energy density of the system.

\begin{figure}
\centering
\includegraphics[width=0.6\textwidth]{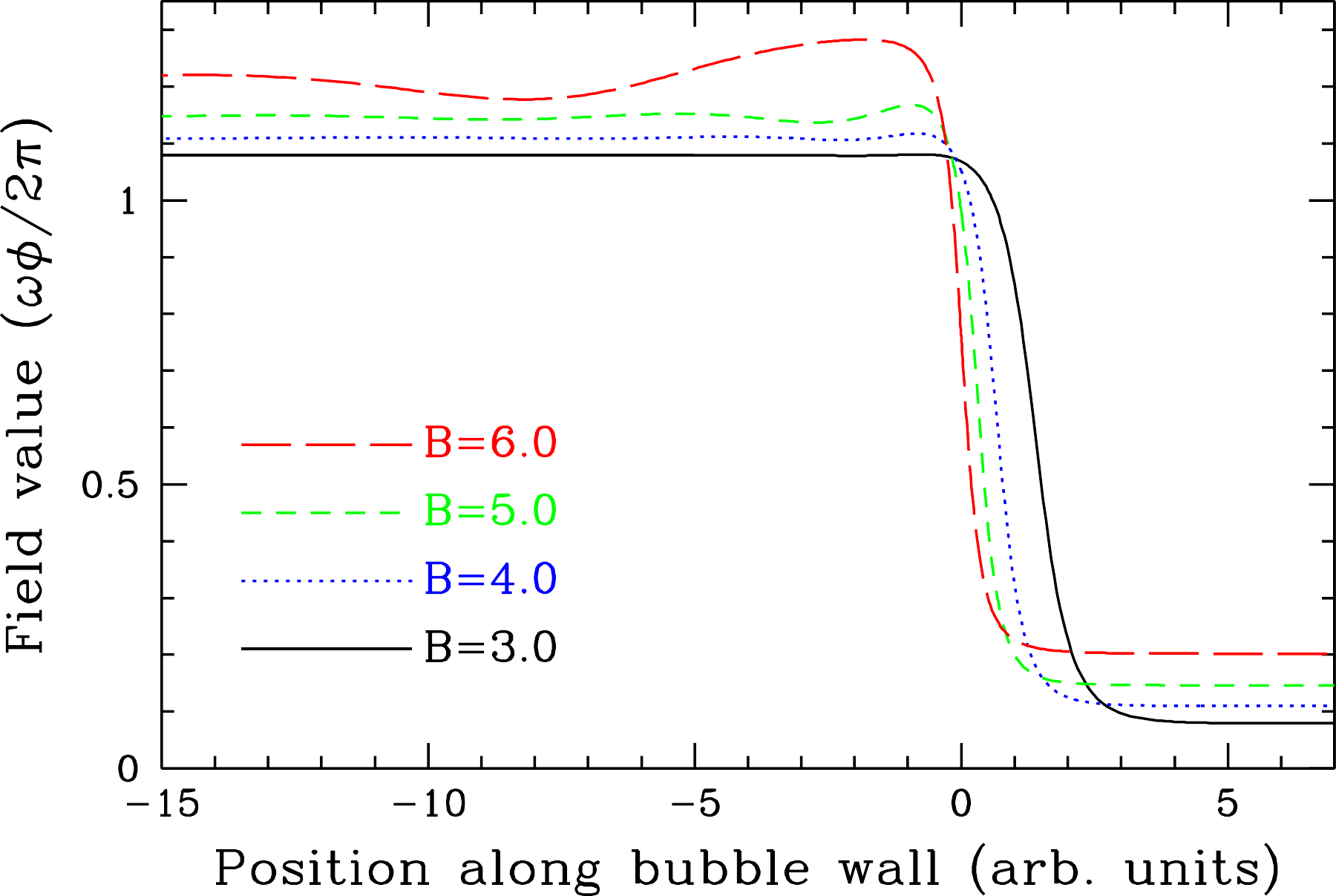}
\caption{\label{ripples}
 profile of a bubble wall, moving towards the right, for several values
 of $B$ the ratio of ``slope'' to ``oscillation'' heights in the
 potential.  The larger the slope, the larger the ``ripples'' which
 appear behind the wall.  For a critical value of $B$ just larger than
 the largest shown, the ripples nucleate the next transition.}
\end{figure}

However, there is another complication of nucleation in a potential with multiple
minima such as that given in Eq.~(\ref{eq:sample-potential}). When the ``slope'' is
relatively large compared to the ``oscillation,'' quantified by the parameter $B
\equiv 2\pi V'_0/V_1\omega$,   not all of the potential energy liberated by a bubble
wall goes into accelerating the wall.  Some goes into creating ripples in the $\varphi$
field, as illustrated in Figure \ref{ripples}, which shows  moving bubble wall
profiles  for $B=3,4,5,6$.  Solving the field equations for a single accelerating bubble wall
using the potential of Eq.~(\ref{eq:sample-potential}), we find that for $B>6.02$
these ripples grow large enough that they overshoot the next barrier and induce the
next phase transition.  This sets off a cascade in which $\varphi$ ``runs away'' down
the potential.  So $B<6.02$ is required to ensure that bubble nucleations only induce
a single phase transition and do not set off a catastrophic run down the potential. 
This puts a lower limit on the potential barrier relative to the potential drop
between minima.  (Note that at $B=2\pi$ the minima disappear; so the
limiting value of $B$ is quite close to where the potential becomes
monotonic.)

This requirement on the potential has consequences for the bubble
nucleation action.  Applying the procedure of Coleman {\it et al.}
\cite{Coleman:1977py, Coleman:1980aw} to find the nucleation action $S$
for the potential in Eq.~(\ref{eq:sample-potential}) (without making any
thin-wall approximation), we find the result
depends only on $B$ and $V_1 \omega^4$:
$S = (V_1 \omega^4)^{-1} F(B)$, where the function $F$ is fit by
$F = (2\pi)^4(348.6/B^3 - 0.738/B^2 - 8.44/B)$ on the interval $1 \lsim B \lsim 6$.  
In particular, at the critical value $B=6.02$, we find that $F = 333.5$.  
(For smaller values of $B$, $F$ is even larger, so this represents the best
case for getting a large tunneling rate.)  
The nucleation action can only be made small 
by choosing a large value for $V_1\omega^4$.  But this quantity
 is also the maximal value of the fourth derivative of the
potential.  Provided that the potential can be regarded as the
effective potential of a canonically normalized scalar field in a
four-dimensional quantum field theory, the potential is only stable
against large loop corrections if $V'''' \sim V_1 \omega^4$ is small
enough,%
\footnote{%
  If we consider a potential with $V'''' \gg 2\pi$, defined at some
  physical scale $\mu_1$, then the potential as measured at a more
  infrared scale $\mu_2$ will have a smaller value 
  $V'''' \lsim 2\pi / \ln(\mu_1/\mu_2)$.  The condition $V'''' \lsim
  2\pi$ is therefore unavoidable if we demand that the theory {\sl exist} at
  an energy scale a few times higher than the mass scale of the
  potential $\sim V_1 \omega^2$.}
say, $V'''' < 2\pi$.
Imposing this condition, we find $S \geq 53$.  This leads to a bubble
nucleation rate $\Gamma \sim \exp(-S)$ which may be too 
small to allow the phase transitions to complete in a Hubble time.  We will see
that this requirement makes it difficult, though not impossible, to find 
consistent models of chain inflation.

\section{Mean evolution of the field}
\label{sec:evolution}

We now proceed to study how the mean field $\langle\varphi\rangle$ evolves
in chain inflation, which is quite different from the slow-roll dynamics
of conventional inflation models.  
We work in comoving coordinate $x$ and conformal time $\tau$. In these
coordinates, light cones are straight lines with slope $45^{\circ}$.
The tunneling rate can be written as a function of time as
\begin{align}
  \Gamma_c = \frac{\Gamma}{a^4} = \frac{\Gamma}{H^4\tau^4}~,
\end{align}
where $\Gamma_c$ is the tunneling rate in comoving coordinates and
conformal time, and $\Gamma$ is the tunneling rate in physical coordinates. We
assume $\Gamma$ is a constant, or varying slowly along the chain of
meta-stable vacua.%
\footnote{%
    If $\Gamma$ varies significantly in one Hubble time, the
    perturbation spectrum will have a large tilt, in contradiction to
    observations.}

To model the phase transition dynamics, a random variable $X(x,\tau)$ is
placed on every spacetime point, denoting the probability density for a
spontaneous (in other words, not from bubble wall collision) 
tunneling event to take place. 
As we discussed, the collision of bubble walls does not stop the
propagation of a bubble wall; so any spontaneous nucleation event along the past
light-cone of some spacetime point $(x,\tau)$ will cause a bubble wall
to induce a phase transition at $(x,\tau)$, and any spontaneous nucleation event
{\sl inside} the past light cone of $(x,\tau)$ has caused such a
transition along the past world-line of an observer at $(x,\tau)$.
Therefore the scalar field value at a point $(x,\tau)$ is
determined by a sum over all spontaneous nucleation events in the past light cone.
We find
\begin{align}\label{eq:phi_rand}
  \varphi(x,\tau) = \Delta\varphi \int^{\tau} d\tau' \int d^3 x' \,
 \Theta[ (\tau-\tau')^2-(x-x')^2]\,  X(x',\tau')~,
\end{align}
where $\Delta\varphi \equiv 2\pi / \omega$ is the distance between
neighboring  minima in field space, and the integral with the Heaviside
$\Theta$ function is over the interior of the
past light cone of $(x,\tau)$. The random variables $\{X(x,\tau)\}$
obey the Poisson distribution
\begin{align}
  X\sim \mathrm{Pois}\left(\Gamma_c ~\mathrm{d}^3x
    \mathrm{d}\tau\right)
  =
  \mathrm{Pois}\left(\frac{\Gamma \mathrm{d}^3x\mathrm{d}\tau}{H^4\tau^4}
  \right)~.
\end{align}

Note that the sum of Poisson random variables is also a Poisson
random variable, with
\begin{align}\label{eq:sum-pois}
  X\sim \mathrm{Pois}(\mu_X)~,\quad Y\sim \mathrm{Pois}(\mu_Y)
\quad \Rightarrow \quad X+Y \sim \mathrm{Pois}(\mu_X+\mu_Y)~.
\end{align}
Thus the summation over the past light cone (analog of $\mu_X+\mu_Y$
in equation (\ref{eq:sum-pois})) takes the form
\begin{align}\label{eq:lambda}
  \int_{\tau_i}^{\tau}\mathrm{d}\eta~
  \frac{4}{3}\pi(\tau-\eta)^3 \frac{\Gamma}{H^4\eta^4}
  =\frac{\varphi_i}{\Delta\varphi}
  +\frac{4\pi\Gamma}{3H^4}\log\left(\frac{\tau_i}{\tau}\right)
  =\frac{\varphi_i}{\Delta\varphi} + \frac{4\pi\Gamma}{3H^3}(t-t_i)~,
\end{align}
where the constant $\varphi_i$ is the field value at an initial time
$\tau_i$, and $t$  denotes proper times of a comoving observer such
that $dt=ad\tau$. Here we have used the approximation $|\tau_i|\gg |\tau|$.
Because of the exponential expansion of space, this approximation
should hold for any reasonably early $\tau_i$. Equation
(\ref{eq:lambda}) means that $\varphi(x,\tau)$ obeys the distribution
\begin{align}
  \varphi(x,\tau) \sim \Delta\varphi \times \mathrm{Pois} \left(\frac{\varphi_i}{\Delta\varphi} +
    \frac{4\pi\Gamma}{3H^3}(t-t_i)\right)~. 
\end{align}
{}From properties of the Poisson distribution ($\langle X\rangle=\mu$
for $X\sim \mathrm{Pois}(\mu)$), the expectation
value of $\varphi(x,\tau)$ is
\begin{align}\label{eq:phi}
  \langle\varphi(x,\tau)\rangle = \varphi_i +
    \frac{4\pi\Gamma\Delta\varphi}{3H^3}(t-t_i)~. 
\end{align}

\section{The power spectrum}
\label{sec:power-spectrum}

\begin{figure}[hptb]
\centering
\includegraphics[width=0.8\textwidth]{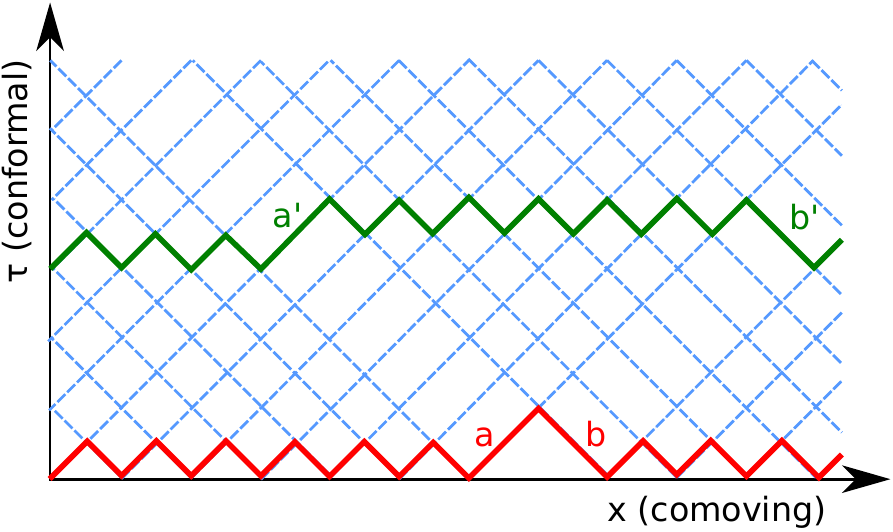}
\caption{\label{fig:prop}The propagation of fluctuations for single
  field chain inflation. The lower heavy (red) line  denotes
  an initial slice. The
  fluctuation on this initial slice is propagated to later slices, 
	for
  example the one denoted by upper heavy (green) line, and becomes
  conserved on super-Hubble scales.}
\end{figure}

In this section, we first calculate the position space two point
function and then Fourier transform to get the power spectrum.
Before doing so, it is helpful to first have an intuitive
understanding of the fluctuations. As illustrated in figure
\ref{fig:prop}, bubble walls move nearly at the speed of light, and the
collision of bubble walls does not affect their dynamics. When there is
a fluctuation at an initial slice (denoted by $a$ and $b$), it
propagates at the speed of light to $a'$ and $b'$ at a
later slice. Since fluctuations cannot propagate very far in the comoving
diagram, they are conserved on super-Hubble scales.

To calculate the position-space two-point correlation
$\langle\delta\varphi(x,\tau)\delta\varphi(x+r,\tau)\rangle$, we note
that $\varphi(x,\tau)$ and $\varphi(x+r,\tau)$ can be expressed in
terms of equation (\ref{eq:phi_rand}). In the overlapping region between the
past light cone of $(\mathbf{x},\tau)$ and $(\mathbf{x+r},\tau)$, the
random variables in the summation for $\varphi(x,\tau)$ and
$\varphi(x+r,\tau)$ are identical. On the other hand, outside the
overlapping region, the random variables in the summation for
$\varphi(x,\tau)$ and $\varphi(x+r,\tau)$ are independent. The latter
results in zero expectation value for correlations, and the former
leads to the connected part of the two-point function
\begin{align}\label{eq:2pt_pos}
  \langle\delta\varphi(x,\tau)\delta\varphi(x+r,\tau)\rangle_c
  = \Delta\varphi^2\int_{-\infty}^{\tau} \mathrm{d}\eta ~ V_2(\eta)
  \frac{\Gamma}{H^4\eta^4}~,
\end{align}
where
\begin{align}
  V_2(\eta) \equiv \left\{
    \begin{array}{ll}
      \frac{4}{3}\pi (\tau-\eta-r/2)^2 (\tau-\eta+r/4)
      & \mathrm{when~} r<2(\tau-\eta)\\
      0 & \mathrm{when~} r\geq 2(\tau-\eta)
    \end{array}
  \right.
\end{align}
is the intersection volume of the two past light cones at time
$\eta$.\footnote{We used the properties $\langle X\rangle = \mu$,
$\langle X^2\rangle = \mu^2 + \mu$ of the distribution Pois$(\mu)$
to derive this result.}\  
 Here we have put the start of chain inflation at
$\tau_i=-\infty$. In other words we assume chain inflation begins
several e-folds before the largest currently observable scales exit the
horizon.  Otherwise there can be modifications to the lowest observable
multipoles, which depend on the details of what happens before chain
inflation, an issue we do not want to address here.

The integration in equation (\ref{eq:2pt_pos})
has an IR divergence, proportional to $\log\tau$. The IR divergence
is a characteristic of random walk behavior in position space, 
which is also
present in standard slow roll inflation. We do not need to worry about
this divergence because it is absent after the Fourier
transformation. The Fourier space two-point correlation function is
given by
\begin{align}\label{eq:2pt_mom}
  \langle \delta\varphi_{\mathbf k_1}(\tau) \delta\varphi_{\mathbf k_1}(\tau)
  \rangle = (2\pi)^3\delta^3(\mathbf{k}_1+\mathbf{k}_2)
  \int_0^\infty \mathrm{d}r \frac{4\pi r \sin(kr)}{k}
  \langle\delta\varphi(0,\tau)\delta\varphi(r,\tau)\rangle~.
\end{align}
Inserting equation (\ref{eq:2pt_pos}) into (\ref{eq:2pt_mom}),
interchanging the order of integration, and then taking the limit $k\equiv
k_1=k_2\ll
-1/\tau$ 
to consider the super-Hubble modes, we obtain
\begin{align}
  \langle \delta\varphi_{\mathbf k_1}(\tau) \delta\varphi_{\mathbf
    k_1}(\tau)\rangle
  = (2\pi)^3 \delta^3
  (\mathbf{k}_1+\mathbf{k}_2)\frac{8\pi^3\Gamma\Delta\varphi^2}{3H^4 k^3}~.
\end{align}
Using the relation
\begin{align}
  \zeta = -\frac{H}{\dot\varphi}\delta\varphi~,
\end{align}
where $\zeta$ is the curvature perturbation in the uniform energy
density slice, and the time derivative of equation (\ref{eq:phi}), we have
\begin{align}
  \langle \zeta_{\mathbf k_1} \zeta_{\mathbf k_2}\rangle = (2\pi)^3
  \delta^3 (\mathbf{k}_1+\mathbf{k}_2) \frac{3\pi H^4}{2\Gamma k^3}~.
\end{align}
Comparing to the relation
\begin{align}
  \langle \zeta_{\mathbf k_1} \zeta_{\mathbf k_2}\rangle = (2\pi)^3
  \delta^3 (\mathbf{k}_1+\mathbf{k}_2) \frac{2\pi^2}{k^3} P_\zeta
\end{align}
gives the normalization of the power spectrum,
\begin{align}
  P_\zeta = \frac{3H^4}{4\pi\Gamma}~.
\end{align}
Inserting the observed value $P_\zeta = 2.42\times 10^{-9}$, 
we find one of our main results, that $\Gamma / H^4 \simeq 10^{8}$.
The number of tunnelings per unit Hubble time and volume has to
be surprisingly large.  

Our result disagrees with previous computations of the normalization
of the power spectrum  \cite{Feldstein:2006hm,Chialva:2008zw}, whose
validity was assumed by  subsequent papers such as ref.\
\cite{Ashoorioon:2008nh}.  Our different finding is in part due to the
fact that previous references did not consider bubble nucleations
triggered by bubble wall collisions.  Our methodologies have further
differences; ref.\ \cite{Feldstein:2006hm} calculated the spectrum
numerically and obtained $P_\zeta \sim (H^4/\Gamma)^{5/3}$  in
contrast to our result $P_\zeta \sim (H^4/\Gamma)$.  Ref.\ 
\cite{Chialva:2008zw} obtained the conventional result $P_\zeta \sim
H^2/(M_p^2\epsilon)$ where $\epsilon = -\dot H/H^2$ is the Hubble slow
roll parameter.  This arises from assuming vacuum initial conditions
to solve the equation of motion for the perturbations, which is not
valid for chain inflation.

The spectral index is
\begin{align}
  n_s -1 = \frac{\mathrm{d}\log P_\zeta }{\mathrm{d}\log k} =
  4\frac{\dot H}{H^2} -\frac{\dot\Gamma}{H\Gamma}~.
\label{nseq}
\end{align}
When $\dot{\Gamma}$ is negligible,  the spectrum
is necessarily red since $\dot H < 0$. However, $\dot\Gamma/H\Gamma$
depends on how $V_0,V_1$ vary with $\varphi$, with indeterminate sign and
magnitude.  Therefore chain
inflation does not make a definite prediction for the spectral tilt;
it can  generate a large tilt of either sign.  In particular, since we
saw that $\Gamma$ is extremely exponentially sensitive to $B$ the ratio
of slope to oscillation height in the potential, the slightest evolution
of $B$ along the potential will significantly change $\Gamma$.  It
appears to require a carefully tuned potential to ensure that the spectral
tilt is {\sl not} large.

\section{Explicit realizations}
\label{sec:models}

In this section we investigate the extent to which one can find 
simple working models of chain inflation, taking into account all the
phenomenological constraints and requirements of consistency.
We take the simplest possible potential of the form (\ref{eq:sample-potential}), with
\begin{align}
	V(\varphi) = V_0 - A\varphi + V_1\sin(\omega\varphi)
\end{align}
where $V_0$ is constant and $A > 0$.  This form is assumed to hold during inflation;
for $\varphi$ exceeding some maximum value $\varphi_{\rm end}$, 
 $V_0$ must become field-dependent and 
reach a minimum in order for inflation to end.  Motivated by the discussion in
section \ref{sec:setup-backgr-dynam}, we will take the optimal value for $B$,
\begin{align}
	B \equiv  {2\pi A\over V_1\omega} = 6.02
\label{Bchoice}
\end{align}
so as to alleviate as much as possible the tension between having a large enough
tunneling rate and a small enough value of the effective self-coupling of the
field,
\begin{align}
	\lambda = {\rm max}\,V'''' = V_1\omega^4
\end{align}
We further assume that $V_0 \gg V_1, |A\varphi|$
so that the Hubble parameter is approximately constant during inflation, $H \cong
\sqrt{V_0/(3M_p^2)}$.

To  estimate the tunneling rate, we  use \cite{Callan:1977pt}
\begin{align}
	\Gamma \sim S^2 M^4 e^{-S}
\label{Gammaeq}
\end{align}
where $S$ is the action of the bounce solution,
and $M$ is the mass scale from the fluctuation determinant, which 
we estimate using the curvature of the potential, $M^2\sim
V_1\omega^2$.   The bounce action was already computed in 
section \ref{sec:setup-backgr-dynam}, $S = 333.5/\lambda$.
Using (\ref{Gammaeq}) and $H \cong \sqrt{V_0/(3M_p^2)}$ we can express the
requirement $\Gamma/H^4=10^8$, coming from the normalization of the
power spectrum, in terms of the model parameters:
\begin{align}
	9 M_p^4 V_0^{-2}\,{V_1^2\omega^4}\, S^2 e^{-S}  = 10^8 ~.
\label{normeq}
\end{align}

Using the observed value of the spectral index, $n_s > 0.97$
\cite{Komatsu:2010fb},
we obtain a constraint from Eq.\ (\ref{nseq}).  Although $\Gamma$ is
strictly constant in the present model, one could imagine that
the parameters of the model vary slowly with $\varphi$ and thus
provide a source of deviation from the Harrison-Zeldovich spectrum
$n_s=1$ through $\dot\Gamma$.  It is however unlikely that the two terms $4\dot H/H^2$
and $\dot\Gamma/H\Gamma$ should be finely tuned against each other 
to give a small result.  We therefore demand that $|4\dot H/H^2| <
0.04$ (the 68\% c.l.\ limit).  This leads to ${A\Delta\varphi/V_0} < 5\times 10^{-11}$.
Using (\ref{Bchoice}) and (\ref{normeq}) to eliminate $A$ and $V_0$, this leads to an upper bound
on $\Delta\varphi = 2\pi/\omega$, depending upon the effective self-coupling $\lambda
= V_1\omega^4$,   
\begin{align}
	\log_{10}(\Delta\varphi / M_p) < -5.25 - \frac{1}{2}\log_{10}\lambda - {36.2\over\lambda}
\label{spectlim}
\end{align}

\begin{figure}[hptb]
\centering
\includegraphics[width=0.8\textwidth]{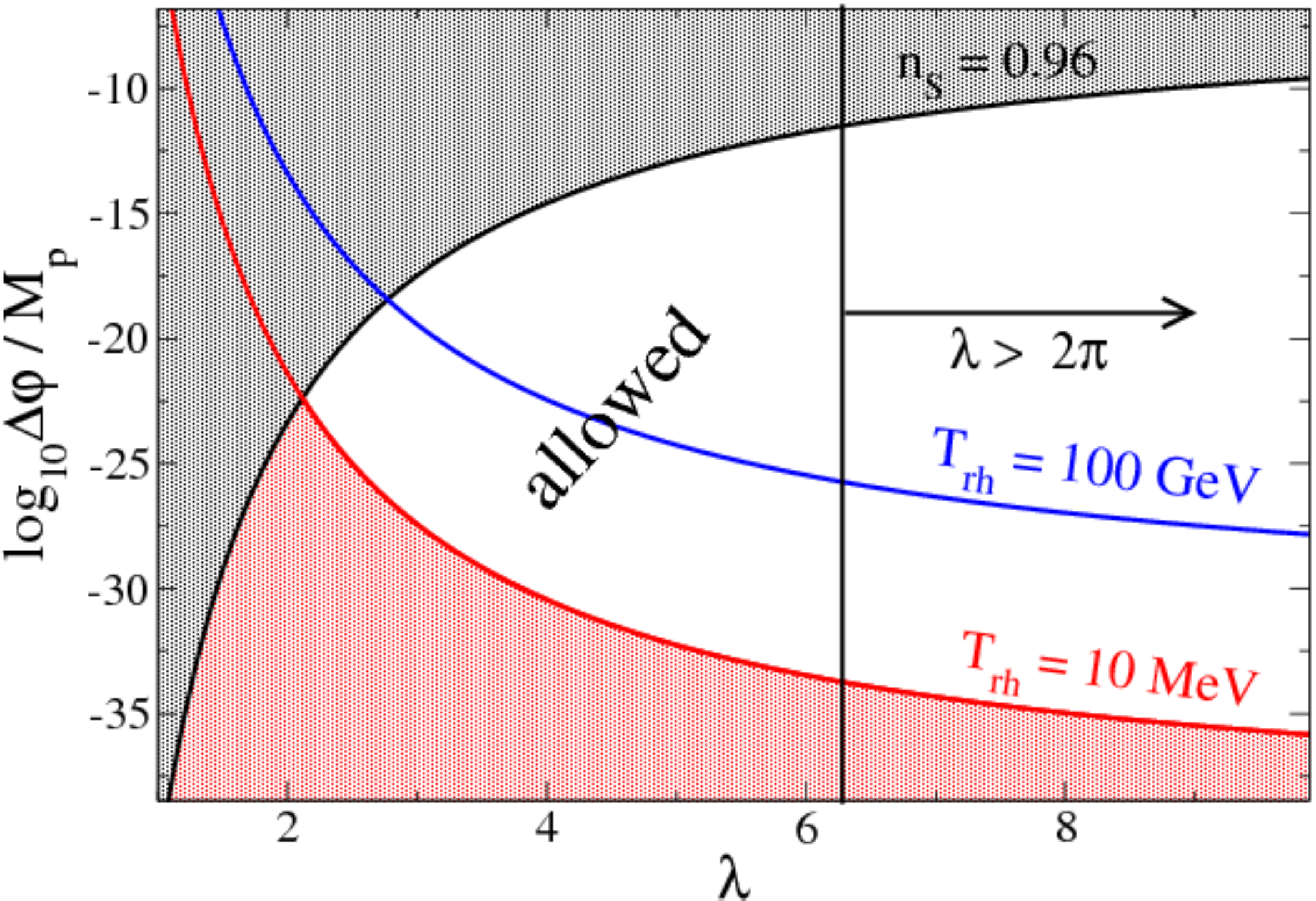}
\caption{\label{fig:const} Unshaded regions in the $\lambda$-$\Delta\varphi$ 
plane
are those for which the spectral index is not too red and the reheat temperature
is above 10 MeV.  The wedge-like region labeled ``allowed'' additionally fulfills the
consistency requirement $\lambda < 2\pi$.}
\end{figure}

For our description in terms of nucleation of individual bubbles to be
consistent, the bubbles must not be strongly overlapping.  For the 
thick-walled bubbles we are considering, the initial radius of the bubble
is of order the inverse mass scale, $r\sim M^{-1} \sim \omega/\sqrt{\lambda}$.
Since $\Gamma/H^4 = 10^8$,
there must be room for $10^8$ such bubbles in a 4-dimensional Hubble
volume: $10^8\times 2\pi^2 r^4 < H^{-4}$.  Using $3M_p^2H^2=V_0$ and (\ref{normeq}) to
eliminate $V_0$, this becomes an upper limit on $\lambda$, 
\begin{align}
	2\pi^2 S^2 e^{-S} < 1 
\end{align}
which for the case $B=6$ is solved by $\lambda < 49$.   This of course is far weaker
than the field theoretic consistency bound $\lambda < 2\pi$ discussed in section
\ref{sec:setup-backgr-dynam}.

One would like to ensure that the change in $\varphi$
remains less than $M_p$ over the course of the field's evolution.
Assuming there were $N_e$ e-foldings of inflation, the total change
in $\varphi$ implied by Eq.\ (\ref{eq:phi}) is
$\Delta^{(N_e)}\varphi = (4\pi\Gamma/ 3H^4)\,\Delta\varphi N_e$
where $\Delta\varphi = 2\pi/\omega$ is the change in $\varphi$ between successive
minima of the potential. We demand that $\Delta^{(N_e)}\varphi < M_p$ for $N_e = 60$.
This gives the upper bound $\Delta\varphi < 10^{-10} M_p$, which is
weaker than (\ref{spectlim}) and thus does not play an important role.

On the other hand, the requirement of a large enough reheating temperature 
$T_{\rm rh}$ does
give an interesting constraint on $\Delta\varphi$ that is complementary to
(\ref{spectlim}).  The reason is that $T_{\rm rh}\sim V_0^{1/4}$ and from
(\ref{normeq}) $V_0 \propto (\Delta\varphi)^2$, so low values of $\Delta\varphi$
as needed to satisfy (\ref{spectlim}) may also lead to excessively low reheat
temperatures.  If we generously assume such efficient reheating that 
$T_{\rm rh}= V_0^{1/4}$ and that $T_{\rm rh} > 10$ MeV so that big bang
nucleosynthesis can proceed normally, we obtain 
\begin{align}
	\log_{10}(\Delta\varphi/M_p) > -39.5 + 2\log_{10}\left({T_{\rm rh}\over 10{\rm\
MeV}}\right)  + {36.2\over\lambda}
\label{Trhlim}
\end{align} 

We have summarized the most important constraints in figure \ref{fig:const}, which
plots them in the $\lambda$-$\Delta\varphi$ plane.  If $T_{\rm rh} = 10$  MeV, there
is an interval of $\lambda\in [2.1,2\pi]$ where it is possible to find consistent
models; at $T_{\rm rh} = 100$ GeV, more preferable for being able to generate a
baryon asymmetry, the range shrinks to $\lambda\in [2.8,2\pi]$  If one is willing to
relax the field-theoretic consistency condition $\lambda < 2 \pi$ for some reason, it
becomes much easier to find viable models.

\section{Non-Gaussianities}
\label{sec:non-gaussianities}

Unlike the case of perturbative quantum fluctuations from the vacuum,
in chain inflation the perturbations come from light-cone shaped
bubbles. Thus one might worry about the Gaussianity of the
scenario. However, as there are a great number of bubbles per Hubble
time and Hubble volume, the Poisson-distributed field fluctuation
approaches a Gaussian distribution. Further, as we shall show, all the
non-Gaussianity from $n$-point correlations turns out to be small.

The three-point correlation function can be calculated similarly to
the case in Section \ref{sec:power-spectrum}. Note that for $X\sim
\mathrm{Pois}(\mu)$, 
$\langle (X-\langle X\rangle)^3 \rangle = \mu$. We have
\begin{align}
  \langle\delta\varphi^3\rangle
  = \Delta\varphi^3 \int_{-\infty}^{\tau} \mathrm{d}\eta ~ V_3(\eta)
  \frac{\Gamma}{H^4\eta^4}~,
\end{align}
where $V_3(\eta)$ is the volume of the intersection region of three balls with
radius $\tau-\eta$. Writing in terms of $\zeta$, and comparing to the
definition of $f_{NL}$ \cite{Komatsu:2001rj},
\begin{align}
  \langle \zeta^3 \rangle 
  = (2\pi)^3\delta^3(\mathbf{k}_1+\mathbf{k}_2+\mathbf{k}_3)
  \frac{3}{5}f_{NL} P_\zeta^2~,
\end{align}
we find $f_{NL}=\mathcal{O}(1)$.

Higher point correlations can be estimated in the same way. For $X\sim
\mathrm{Pois}(\mu)$, $\langle (X-\langle X\rangle)^4 \rangle = \mu+3\mu^2$,
$\langle (X-\langle X \rangle)^5\rangle = \mu + 10 \mu^2$,
$\langle (X-\langle X\rangle)^6 \rangle =
\mu + 25\mu^2+15\mu^3$. For the four-point function, 
the $3\mu^3$ term must be subtracted  to remove
the disconnected part. Thus the connected part has $\langle
(X-\langle X\rangle)^4
\rangle_c = \mu$.  In the language of non-Gaussian estimators,
this corresponds to $g_{NL}=\mathcal{O}(1)$. Similarly, the five- and
six-point estimators $h_{NL}$ and $i_{NL}$ defined in
\cite{Lin:2010ua} are also of order one. This statement generalizes
to arbitrary $n$-point correlation functions. Thus
non-Gaussianity for single-field chain inflation is small at any
order.

\section{Generalizations}
\label{sec:generalizations}

In this section we briefly discuss three generalizations of single-field
chain inflation, namely the curvaton scenario, the multi-field
case, and a noncanonical kinetic term.

We start with the curvaton scenario, where there is a second
field whose rolling (or other dynamics) dominates the change
in the total energy density. In this case, the fluctuation in the
chain tunneling direction (still denoted by $\varphi$ here) is an
isocurvature perturbation. This isocurvature perturbation converts to a
curvature perturbation later by oscillation of $\varphi$ around its
minimum, or if the chain of tunnelings ends after the
reheating associated with the inflaton. Suppose there is no direct
interaction (except via gravity) between the inflaton sector and
the $\varphi$ field. Then the component curvature perturbation
$\zeta_\varphi$ is conserved, which is related to the total curvature
perturbation $\zeta$ as
\begin{align}
  \zeta = r \zeta_\varphi~,\qquad r = \frac{\rho_\varphi+p_\varphi}{\rho+p}~,
\end{align}
where $\rho$, $p$, $\rho_\varphi$, $p_\varphi$ are the total energy
density and pressure and the energy density and pressure for the
$\varphi$ field respectively. $\zeta_\varphi$ is related to the field
fluctuation as $\zeta_\varphi = -H\delta\varphi/\dot\varphi$. After 
the
curvaton $\varphi$ decays, $r$ becomes constant, thus $\zeta$ is
subsequently conserved. The power spectrum can be calculated similarly
to the last section as
\begin{align}
  P_\zeta = \frac{3r^2H^4}{4\pi\Gamma}~.
\end{align}
Since it is suppressed by $r^2$, the original fluctuation in the
$\varphi$ field may be large.  This reduces the number of minima needed
in the $\varphi$ potential, but at the same time it increases
the non-Gaussianity
$f_{NL}=\mathcal{O}(1/r)$, which should be the best
estimator for non-Gaussianity, and is potentially observable. 

It is also interesting to generalize the simplest scenario to a
multi-field case, where a chain of tunnelings can take place in more
than one field direction. Whenever successive tunnelings occur in
orthogonal field-space directions, the collision of bubble walls will
not trigger a new phase nucleation.

In this case, for a given nucleation rate $\Gamma$, an individual
observer will undergo fewer phase transitions.  Specifically, the mean
inter-bubble spacing is $\delta r \sim \Gamma^{-1/4}$, the mean time
between phase changes for a comoving observer is $\delta t \sim \delta
r$, and so the mean number of phase changes per Hubble time is
$\sim (\Gamma /H^4)^{1/4}$, not $\sim \Gamma/H^4$ as we found for the
single-field case.  

\begin{figure}[hptb]
\centering
\includegraphics[width=0.8\textwidth]{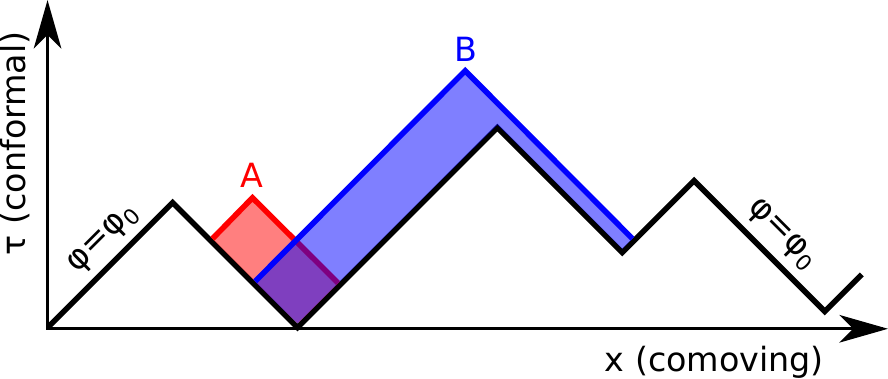}
\caption{\label{fig:multi} The probability for $B$ in the next step of
the chain is larger than that of $A$, even if bubble wall collision no
longer triggers a new bubble nucleation. }
\end{figure}

Perturbations in this scenario remain correlated on scales shorter than
the Hubble scale.  The reason is that, while individual bubble walls no
longer propagate Hubble scale distances, {\sl information} about bubble
nucleations still can.  As an illustration, consider figure
\ref{fig:multi}.  In the figure, one region of space (near $A$) reaches
$\varphi=\varphi_0$ before another region (near $B$).  But the probability
that $B$ will encounter another phase change and get to $\varphi_0+\Delta\varphi$ is
determined by the likelihood of a nucleation occurring in the shaded
area in its past light cone.  This region is larger, and hence the
chance of a nucleation is larger,
than the shaded area in $A$'s past light cone.  Therefore, unevennesses
in the phase boundary allows regions which are ``behind'' to ``catch
up,'' at least on sub-Hubble scales.  This means that perturbations from
the stochastic nature of nucleations get averaged over roughly
Hubble-scale volumes.

However, on super-Hubble scales no such communication is possible.
Therefore the progress of the field $\varphi$ on super-Hubble scales is
still determined by the independent, Poisson-distributed number of
nucleations in the past light cone.  Hence we estimate that the power
spectrum is again
$P_\zeta = (H^4/\Gamma)$, similar to the power spectrum for single-field
chain inflation.  We have not been able to evaluate the order-1
coefficient analytically in this case.  But this is enough to show that
the tunneling rate must be comparable to that in the single-field case,
though the number of minima in the potential can be substantially
smaller ($\sim 10^4$ rather than $\sim 10^{10}$).

A third possible generalization of chain inflation is to allow for 
a noncanonical kinetic term.  There is some motivation to do so since
ref.\ \cite{Brown:2007zzh}
pointed out that the tunneling rate in a generic scalar field theory with
false minima
can be increased using the
string-motivated DBI action with Lagrangian
\begin{align}
{\cal L} = 
-{1\over f(\varphi)}\left(\sqrt{1 - (\partial\varphi)^2\,f(\varphi)}-1\right)
-V(\varphi) \,.
\end{align} 
If $f$ is a constant, the tunneling action can be reduced by at most a factor of 
order unity; to achieve a greater reduction, one must choose $f$ to be
proportional to the term $V_1\sin(\omega\varphi)$ that gives the
barrier, and tune $f V_1\to 2$.   $S$ is then reduced by the factor
$(1 - fV_1/2)^2$.  However taking $f\propto V_1\sin(\omega\varphi)$ seems quite
contrived, so we consider to what extent the modest reduction in $S$ afforded by
constant $f$ can help to
alleviate the tension between various constraints 
discussed in section \ref{sec:models}.  

It is amusing to notice that the tunneling action $S$ can still be computed exactly
within the thin-wall approximation (even though we did not use this approximation  in
our previous analysis) for the model of section \ref{sec:models} and the DBI action
with constant $f$.  In the thin-wall approximation, $S$ is reduced relative to
its usual value by the factor
	$\frac12\left(1 + {{1-a}\over 2\sqrt{a}}\ln{1+\sqrt{a}\over
	1-\sqrt{a}}\right)$
where $a = fV_1/2$.  When $a=0$ we recover the usual result, and when $a=1$ we obtain
the maximum possible reduction in $S$ for constant $f$, a factor of $1/2$.  Let us
take this as indicative of how well one might hope to do in the case of interest where
the thin-wall approximation is not valid.  (We are not enthusiastic enough about this
generalization to do a full numerical solution of the bubbles.)  The most important
constraints, those on the spectral index and the reheat temperature, 
are both relaxed since the spectral index bound goes like
$\Delta\varphi < \#\ S^{1/2}
e^{-S/4}$ while the reheating bound has the form $\Delta\varphi > \#\ 
S^{-1/2} e^{S/4}$.  We find that the minimum allowed value of $\lambda$ decreases to
$1.1$ for $T_{\rm rh}=10$ MeV, and to $1.4$ for $T_{\rm rh}=100$ GeV.  The model
remains rather strongly coupled, but less so than for a standard kinetic term.

\section{Conclusion}
\label{sec:conclusion}

To conclude, we reconsidered models of chain inflation focusing on the
single-field case, and giving what we consider to be the first correct
derivation of the spectrum of fluctuations, taking into account their
Poisson nature as a result of the dynamics of bubble collisions.  We
showed that the power spectrum can be calculated analytically and that
consistency with the size of density perturbations measured today
requires $10^8$ tunnelings per Hubble volume per e-fold. Therefore
single-field models of chain inflation need a potential with at least
$O(60)\times 10^8 \sim 10^{10}$ successive minima, a significant
challenge for model building. The spectral tilt depends on details of
the potential; it can be either red or blue and need not be small.
Non-Gaussianities in the model are small, of order unity. Thermal
contributions to the fluctuations are negligible in this scenario
because the energy stored in bubble walls remains in them after they
collide.

A challenge for chain inflation is posed by a standard
requirement of consistency for any field theory. Under
renormalization, couplings run, and large couplings run very fast.  In
order to have any nonnegligible range of scales over which the theory
remains valid, the effective coupling given by the fourth derivative
of the potential must not be much larger than $O(1)$.  We have shown
that this requirement, together with constraints on the spectral index and reheating
temperature, leaves only a small interval of parameter space that gives 
a large enough nucleation rate.

On the other hand, the method used to calculate the power spectrum and
non-Gaussianity in this model may find application for cases where other
discrete processes instead of a continuous process take over the
role of quantum fluctuations during inflation -- for example, if the
scale of inflation is very high and physics becomes discrete near the
Planck scale.

\section*{Acknowledgment}
We thank Bret Underwood for discussion. YW is supported by grants from
McGill University, Fonds Qu\'eb\'ecois de la Recherche sur la Nature et
les Technologies (FQRNT), the Institute of Particle Physics (Canada)
and the Foundational Questions Institute.
This work was supported in part
by the Natural Sciences and Engineering Research Council of Canada.

\end{document}